\documentclass[12pt,preprint]{emulateapj}

\shorttitle{Low-mass high-velocity WD}
\shortauthors{Kawka et al.}

\begin{document}

\title{LP~400-22, a Very Low-mass and High-velocity White Dwarf} 

\author{Adela Kawka}
\affil{Astronomick\'y \'ustav, AV \v{C}R, Fri\v{c}ova 298, 25165 Ond\v{r}ejov, 
Czech Republic}
\email{kawka@sunstel.asu.cas.cz}

\author{St\'ephane Vennes, Terry D. Oswalt, J. Allyn Smith\altaffilmark{1}}
\affil{Department of Physics and Space Sciences, Florida Institute of 
Technology,\\ 150 W University Blvd., Melbourne, FL32901-6975}
\email{svennes@fit.edu, toswalt@fit.edu, jasmith@astro.fit.edu}

\altaffiltext{1}{Visiting Astronomer, KPNO/NOAO,
which is operated by AURA,
under cooperative
agreement with the NSF.}
\and

\author{Nicole M. Silvestri}
\affil{Department of Astronomy, University of Washington, Box 351580, Seattle, 
WA 98195}
\email{nms@astro.washington.edu}

\begin{abstract}
We report the identification of LP~400-22 (WD~2234+222) as a very low-mass and high-velocity white dwarf. 
The ultraviolet {\it GALEX} and optical photometric colors and a spectral line 
analysis of LP~400-22 show this star to have an effective 
temperature of $11\,080\pm140$ K and a surface gravity of $\log{g} = 6.32\pm0.08$.
Therefore, this is a helium core white dwarf with a mass of $0.17\ M_\odot$.
The tangential velocity of this white dwarf is $414\pm43$ km s$^{-1}$, making
it one of the fastest moving white dwarfs known. We discuss probable evolutionary
scenarios for this remarkable object.
\end{abstract}


\section{Introduction}

The vast majority of white dwarfs evolve from normal main-sequence stars 
following normal evolutionary processes. However, ultramassive ($>1.1M_\odot$) 
and inframassive ($<0.40M_\odot$) white dwarfs require special evolutionary 
paths. The formation of low-mass helium white dwarfs 
($M_{\rm WD} \la 0.4M_\odot$) has been shown to be the result of close binary 
evolution \citep[][and references therein]{ibe1986}. Indeed, the Galaxy
is not old enough for these objects to have formed through single star
evolution. The general evolutionary scenario for the formation of low-mass 
helium white dwarfs is that the companion stripped the white dwarf of its
envelope before completing its red giant evolution \citep{kip1967}.

Recently, several very low mass white dwarfs ($M_{\rm WD} \la 0.2 M_\odot$) 
have been discovered as companions to pulsars \citep{van2005}. The orbital 
periods vary from a few hours to several years. The masses of some of
these white dwarfs may be determined from the Shapiro delay of radio 
pulses provided that the system is nearly edge on \citep{loh2005}. 
For example, \citet{jac2003} deduced a mass of $0.20\ M_\odot$ for the 
companion of PSR J1909$-$3744, and obtained a spectrum which confirmed, at 
least qualitatively, the presence of a low mass DA white dwarf.
In addition, the masses of the companions to PSR J1012$+$5307 and 
PSR J1911$-$5958 were determined spectroscopically to be $0.16\ M_\odot$ 
\citep{van1996,cal1998} and $0.18 M_\odot$ \citep{bas2006}, respectively. 
Finally, several low-mass white dwarf candidates were found in the Sloan 
Digital Sky Survey \citep{kle2004}. \citet{lie2004} analyzed the brightest 
star in the sample, SDSS~J123410.37$-$022802.9 and showed that it has a mass 
of $\sim 0.18\ M_\odot$ and that it does not have an obvious neutron 
star companion. 

In this paper, we report the identification of a high velocity white dwarf 
with a very low mass, LP~400-22 (WD2234+222\footnote{Online at http://www.astronomy.villanova.edu/WDCatalog/index.html}, NLTT~54331). Our photometric and 
spectroscopic observations are presented in \S 2.1 and 2.2 respectively. We 
derive the stellar parameters in \S 3 and discuss our results in \S 4.

\section{Observations}

LP~400-22 was spectroscopically identified as a white dwarf as part of a
survey of common-proper motion binaries with suspected white dwarf components 
\citep{osw1993}. We obtained additional high-resolution optical spectra \citep{sil2002} 
as well as new optical photometry \citep{smi1997} as part of the 
same project. More recently, LP~400-22 was observed during the 
{\it Galaxy Evolution Explorer} ({\it GALEX}) all-sky survey.

\subsection{Photometry}

The $BVRI$ photometry for LP~400-21/22 were obtained with the 2.1m
telescope at KPNO on 1995 July 5 UT. A Tek1K CCD (with 
$24\mu$m pixels) operating at the Cassegrain focus was used, providing 
$0\farcs305$ per pixel and a $5\farcm2$ field of view. The data for 
LP~400-21/22 were obtained under photometric conditions.
Standard stars
for this program were chosen from \citet{lan1992}.

We obtained ultraviolet (UV) photometry from  {\it GALEX}
All-Sky Survey\footnote{Available from http://galex.stsci.edu/GR1/}. 
{\it GALEX} provides photometry in two bands, FUV and NUV, which are based on 
the AB system \citep{mor2005,oke1983}. The bandwidth of FUV is
1344 to 1786 \AA\ with an effective wavelength of 1528 \AA. The bandwidth
of NUV is 1771 to 2831 \AA\ with an effective wavelength of 2271 \AA.

\begin{figure}
\plotone{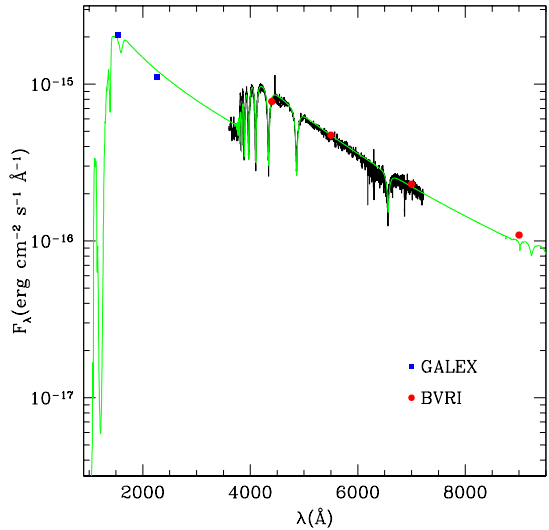}
\caption{The energy distribution of LP~400-22 combining all available data compared to our H-rich model
spectrum at $T_{\rm eff} = 11\,000$ K and $\log{g} = 6.50$ (see \S 3). 
\label{fig_energy_dist}}
\end{figure}

\begin{deluxetable}{ccc}
\tabletypesize{\scriptsize}
\tablecaption{Photometry \label{tbl_phot}}
\tablewidth{0pt}
\tablehead{
\colhead{Band} & \colhead{LP~400-22}& \colhead{LP~400-21}
}
\startdata
FUV\tablenotemark{a} & $18.38\pm0.09$ mag & \nodata \\
    & $18.18\pm0.08$ mag & \nodata \\
NUV\tablenotemark{a} & $18.19\pm0.05$ mag & \nodata \\
    & $18.14\pm0.04$ mag & \nodata \\
B   & $17.338\pm0.025$ mag & $18.742\pm0.025$ mag \\
V   & $17.219\pm0.021$ mag & $17.177\pm0.021$ mag \\
R   & $17.202\pm0.023$ mag & $15.933\pm0.023$ mag \\
I   & $17.210\pm0.024$ mag & $14.340\pm0.023$ mag \\
\enddata
\tablenotetext{a}{The mean of these values are used in this paper.}
\end{deluxetable}

Table~\ref{tbl_phot} presents the optical and ultraviolet photometry and 
Figure~\ref{fig_energy_dist} shows the energy distribution compared to
a synthetic spectrum. 

\subsection{Spectroscopy}

We obtained a low-resolution spectrum of LP~400-22 using the R-C 
spectrograph attached to the 4 m telescope at Kitt Peak National Observatory
(KPNO) on 1988 October 6. The BL250 grating (158 lines/mm) was used to obtain 
a spectral range of 3500 to 6200 \AA\ with a dispersion of 5.52 \AA\ per
pixel and a resolution of 14 \AA.

LP~400-22 was re-observed using the Dual Imaging Spectrogram (DIS) attached to
the 3.5 m telescope at the Apache Point Observatory (APO) on 2001 July 10 and
October 14. The 1200 lines/mm grating was used to obtain a 
spectral range of 3800 to 4600 \AA\ with a dispersion of 1.6 \AA\ per pixel, 
and the 830.8 lines/mm grating was used to obtain a spectral range of 6180 to 
7210 \AA\ with a dispersion of 1.3 \AA\ per pixel. 
A $1\farcs5$ slit was 
used to obtain a spectral resolution of $\sim 2$ \AA\ in 
the blue and $\sim 2.6$ \AA\ in the red.

\section{Determining the Parameters}

In our analysis of LP~400-22, we used a grid of computed pure hydrogen LTE
plane parallel models (see \cite{kaw2006} and references therein for details).
The grid of models extend from $T_{\rm eff} = 7000$ to 16\,000 K (in steps of
1000 K), from 18\,000 to 32\,000 K (in steps of 2000 K) and from 36\,000 to 84\,000 K
(in steps of 4000 K) at $\log{g} = 6.0$ to 9.5 (in steps of 0.25 dex).
All our $\log{g}$ values are in cgs.
We also prepared corresponding grids of synthetic spectra, one of which
includes the effect of Ly$\alpha$ satellites \citep{all1992},
and the other excludes that
effect.

\subsection{Photometry}

\begin{figure*}
\epsscale{0.8}
\plottwo{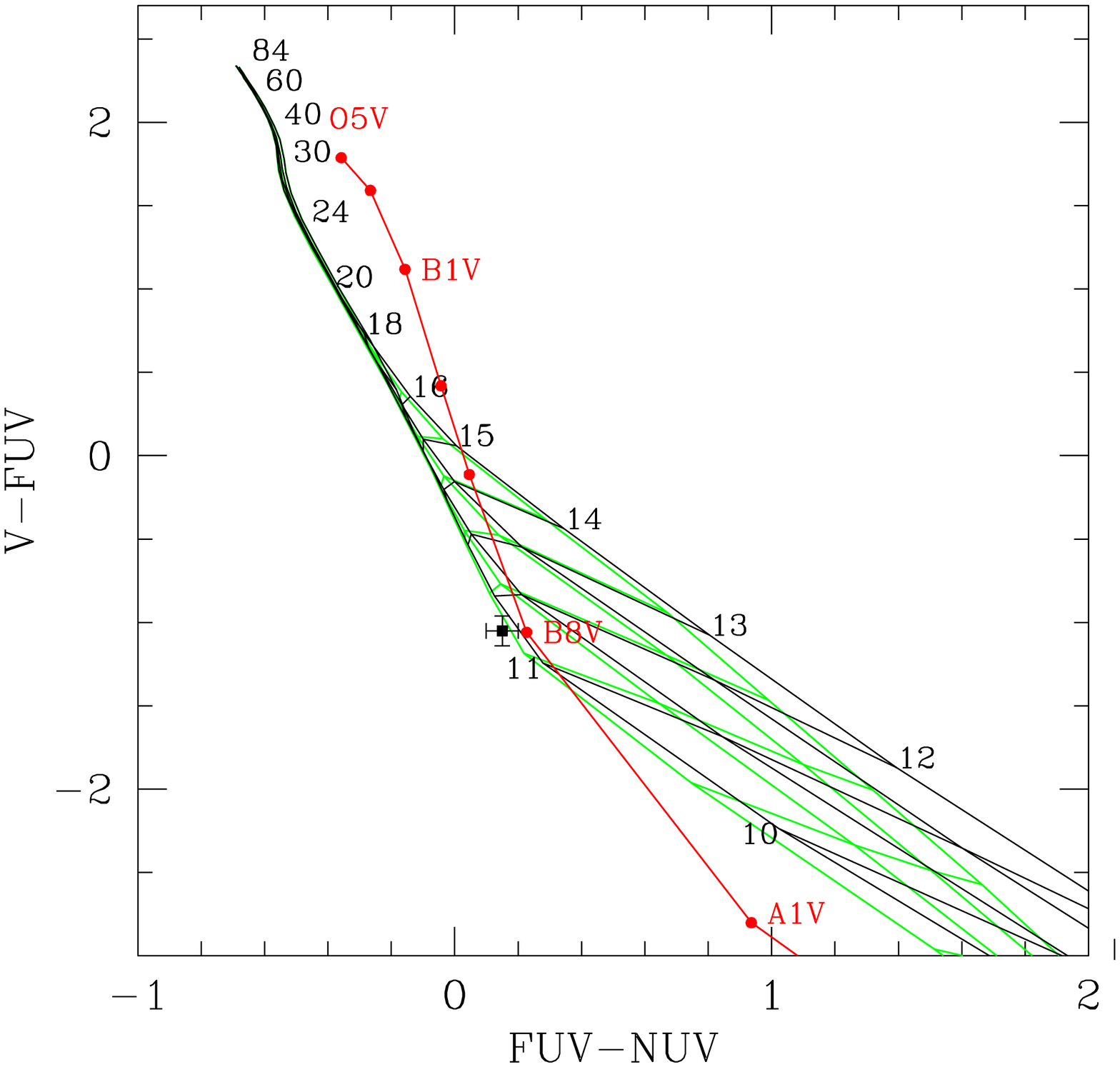}{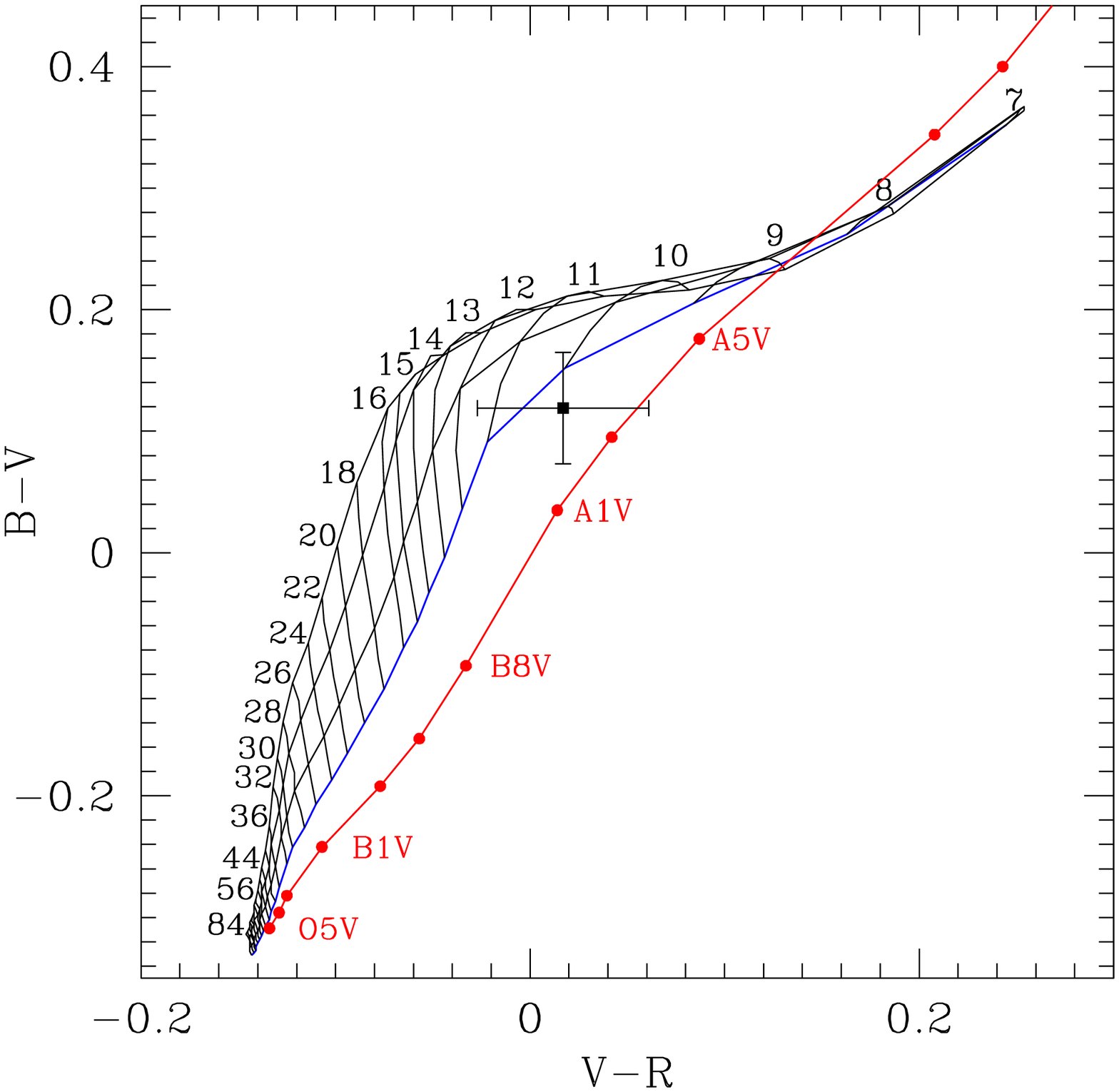}
\caption{{\it GALEX} V-FUV versus FUV-NUV ({\it left}) and optical B-V versus V-R 
({\it right}) diagrams showing
the position of LP~400-22 compared to the DA white dwarf sequence and the
main-sequence (in red). The effective temperatures are indicated in units of 
1000 K and the $\log{g} = 6.0,\ 7.0,\ 8.0$ and $9.0$ (from bottom to top). In 
the optical diagram $\log{g} = 6.0$ is indicated by the blue line. In the 
UV-optical diagram, the grid shown in black includes the Ly$\alpha$ satellite 
and the grid in green does not. 
\label{fig_vmf_fmn}}
\end{figure*}

Using our spectral grid, we have calculated synthetic optical ($BVRI$) and
ultraviolet ($FUV/NUV$) colors. Figure~\ref{fig_vmf_fmn} shows the observed
photometric colors ($V-FUV$ versus $FUV-NUV$ and $B-V$ versus $V-R$) of 
LP~400-22 compared to synthetic white dwarf and main-sequence colors.
We used Kurucz synthetic spectra \citep{kur1993} to calculate our main-sequence 
colors. 

In the UV-optical diagram ($V-FUV$/$FUV-NUV$) of 
Figure~\ref{fig_vmf_fmn} we show two sets of
WD synthetic colors. The grid shown in black includes the effect of Ly$\alpha$
satellites \citep{all1992} as compared to the grid in green which excludes them. 
A comparison of the two grids shows the significant effect that the 
Ly$\alpha$ satellites have on the UV colors at $T_{\rm eff} < 13\,000$ K. 
Comparing the UV-optical photometry of LP~400-22 to the white dwarf grid, 
a low surface gravity $\log{g} \sim 6$ and an effective temperature of 
$\sim 11\,000$ K is implied. 
The optical diagram ($B-V$/$V-R$) in Figure~\ref{fig_vmf_fmn} confirms 
the white dwarf temperature 
of 11\,000 K and the low surface gravity. 

However, when comparing the 
photometry to main-sequence colors, a A3V spectral type is implied in the 
optical and a B8V spectral type in the ultraviolet. Therefore,
the data are incompatible with main-sequence colors. 
Optical and UV colors are useful to distinguish 
white dwarfs from main-sequence stars.

\subsection{Spectroscopy}

\begin{figure}
\epsscale{0.8}
\plotone{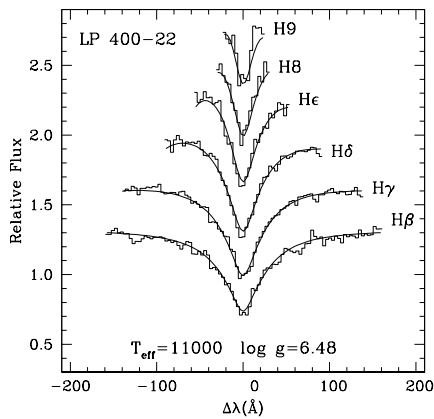}
\caption{Spectral fit of the Balmer lines (H$\beta$ to H9) of the KPNO spectrum
of LP~400-22. \label{fig_specfit_WD2234+222}}
\end{figure}

The Balmer lines of LP~400-22 were analyzed in all three available spectra
using a $\chi^2$ minimization technique. The quoted uncertainties are 
statistical only ($1\sigma$). The Balmer lines (H$\beta$ to H9) in 
the KPNO spectrum were fitted with model spectra which were smoothed to the 
instrumental resolution of 14 \AA, to obtain $T_{\rm eff} = 11\,000\pm350$ K and 
$\log{g} = 6.48\pm0.27$. For the two high-resolution APO spectra we fitted 
H$\alpha$ and H$\gamma$ to H9 with model spectra, to obtain 
$T_{\rm eff} = 11\,060\pm180$ K and $11\,160\pm250$ K, and $\log{g} = 6.46\pm0.13$ 
and $6.22\pm0.10$. The synthetic spectra used in the analysis of the APO
spectra were smoothed with a gaussian profile to the instrumental resolution 
of 2 \AA. Note that the disprepancy in the surface gravities from the 2 APO 
spectra are most likely the result of uncertainties in the flux calibration
around the higher Balmer lines. The Balmer line fit of the KPNO spectrum is 
shown in Figure~\ref{fig_specfit_WD2234+222}. These measurements clearly 
confirm that LP~400-22 is a white dwarf with a low surface gravity.
The calculated weighted average of the temperature and surface gravity is
$T_{\rm eff} = 11\,080\pm140$ K and $\log{g} = 6.32\pm0.08$. 

We used the evolutionary tracks for helium-core white dwarfs of \citet{alt2001} 
and \citet{ser2001} to determine a mass of $0.17\pm0.01\ M_\odot$ and
a cooling age of $5\pm1\times 10^8$ years. Note that the cooling age of the
white dwarf is sensitive to the mass of the hydrogen envelope left before
entering the final cooling track \citep[see][and references therein]{alt2001}.
Residual H-burning in a thick H-envelope causes the white dwarf to cool 
slower as compared to a white dwarf with a thin H-envelope.

The temperature of 11\,080 K places LP~400-22 near the blue edge of the ZZ Ceti 
instability strip \citep{gia2005}. Given the lack of time coverage in our data, we cannot state
whether the star is variable or not. 
Time-series photometry is required to explore variability and place constraints
on the blue edge of the instability strip at the low-mass range.

\section{Discussion}

LP~400-22 was listed in the New Luyten Two-Tenths (NLTT) catalog
\citep{luy1979} to have a common proper motion companion (LP~400-21) 
$338\arcsec$ away. Recently, \citet{sal2003} have revised the coordinates and
proper motions of most stars in the NLTT catalog by cross-correlating the
catalog with the Two Micron All Sky Survey (2MASS) and the USNO-A catalogs.
However LP~400-22 was not detected in 2MASS, and therefore they relisted
Luyten's measurement of the proper motion.
They listed the proper motion of LP~400-22 to be 
$\mu_{\alpha} = 0\farcs1950\pm0\farcs0200$ yr$^{-1}$ and 
$\mu_{\delta} = 0\farcs0563\pm0\farcs0200$ yr$^{-1}$. For LP~400-21 they measured
a proper motion of $\mu_{\alpha} = 0\farcs2158\pm0\farcs0055$ yr$^{-1}$ and 
$\mu_{\delta} = 0\farcs0283\pm0\farcs0055$ yr$^{-1}$. These proper motion 
measurements agree within $2\sigma$ and on this basis the two stars appear to be
a common proper motion binary.
However, similar measurements were reported by
\citet{lep2005}, i.e., $\mu_{\alpha} = 0\farcs198$ yr$^{-1}$ and 
$\mu_{\delta} = 0\farcs053$ yr$^{-1}$ for LP~400-22, and 
$\mu_{\alpha} = 0\farcs228$ yr$^{-1}$ $\mu_{\delta} = 0\farcs020$ yr$^{-1}$ 
for LP~400-21. The quoted uncertainties in the \citet{lep2005} measurements are
$\sim 0\farcs007$ yr$^{-1}$ and, therefore, the diverging proper motions of 
the two stars appears to exclude a physical association. Another way to check
whether the stars are a physical binary is to determine the distance of
each star.

To estimate the distance, we calculated an absolute magnitude of 
$M_V = 9.1\pm0.2$ mag for LP~400-22 and a distance modulus
$(V - M_V) = 8.2$ mag. This places the white dwarf at a distance of 
$430\pm45$ pc. Note that the Galactic extinction for this object is low
and its effect was not included. 
\citet{sil2005} classified LP~400-21 a dM4.5e, and using 
the $M_V/V-I$ relation from \citet{rei1997} we estimate the absolute magnitude 
of the red dwarf as 12.7 mag. The apparent
V magnitude for LP~400-21 is $V=17.177\pm0.021$ mag, and, therefore, the red dwarf is
at a distance of $\sim 80$ pc. \citet{rei1997} note a scatter of values about 
the relation with $\sigma = 0.46$. Even if we consider LP~400-21 at the
extrema of this dispersion, it would place it at a distance of only 
$\sim 100$ pc. The large distance discrepancy makes LP~400-22/21
a coincidental pair rather then a wide binary as has been thought based
on their proper motion alone.

The large distance and high-proper motion of LP~400-22 imply a
large tangential velocity of $414\pm43$ km s$^{-1}$. Only a few white dwarfs
are known to have $v_{\rm tan} > 350$ km s$^{-1}$, with most of these
having halo space velocities \citep{ber2005}. In order to obtain the $U,V,W$ 
space velocity components for LP~400-22, we measured the radial 
velocity of the white dwarf using H$\alpha$ in the APO high-dispersion spectra 
to obtain a heliocentric value of $-50\pm20$ km s$^{-1}$, which is different than the
velocity measured for the red dwarf (7.3 km s$^{-1}$) by \citet{sil2002b}. 
We calculated $U,V,W$ for LP~400-22 using \citet{joh1987} to obtain
$U = -388\pm43,\ V=-81\pm22,\ W=-83\pm22$ km s$^{-1}$.
These velocity components do not agree with either
disk or halo populations \citep{chi2000} and suggest a different origin for 
its peculiar motion. The Galactic orbit for LP~400-22 should be calculated.

Most white dwarfs with $M<0.2\ M_\odot$ are companions to pulsars. We
searched for radio sources in the vicinity of LP~400-22 using 
{\it VizieR}\footnote{http://vizier.u-strasbg.fr/viz-bin/VizieR}, and the 
nearest was that of the galaxy KUG2234+223
$\sim7.5\arcmin$ away. Therefore, if LP~400-22 is a companion to a neutron
star, then it is probably a dead pulsar. Another possibility is that
LP~400-22 has a very cool companion, which should be detectable as 
infrared excess. However, LP~400-22 was not detected by 2MASS.
The two high-dispersion velocity measurements agree
within error bars. However, a series of radial velocity measurements should be 
obtained to establish whether or not LP~400-22 is in a close binary system. 

Yet another possibility for the origin of LP~400-22 is that it may have once
been in a close double-degenerate binary, where the companion has gone through 
a supernova event that disrupted the binary losing the remnant of the donor 
star with a high-velocity and a low mass \citep{han2003}.

\section{Summary}

\begin{deluxetable}{llc}
\tabletypesize{\scriptsize}
\tablecaption{LP~400-22 Parameters \label{tbl_par}}
\tablewidth{0pt}
\tablehead{
\colhead{Parameter} & \colhead{Measurement} & \colhead{Reference}
}
\startdata
Effective Temperature & $11\,080\pm140$ K  & 1 \\
Surface Gravity       & $6.32\pm0.08$    & 1 \\
Mass                  & $0.17\pm0.01 M_\odot$ & 1 \\
$M_V$                 & $9.1\pm0.2$ mag & 1 \\
Distance              & $430\pm45$ pc    & 1 \\
Proper Motion         & $\mu = 0\farcs203$ yr$^{-1}$,$0\farcs205$ yr$^{-1}$ & 2,3 \\
                      & $\theta = 73.9^\circ,\ 75.0^\circ$ & 2,3 \\
Kinematics            & $U = -388\pm43$ km s$^{-1}$ & 1 \\
                      & $V = -81\pm22$ km s$^{-1}$  & 1 \\
                      & $W = -83\pm22$ km s$^{-1}$ & 1 \\
\enddata
\tablerefs{
(1) This work; (2) \citet{luy1979,sal2003}; (3) \citet{lep2005}
}
\end{deluxetable}

We have demonstrated that LP~400-22 is a high-velocity white dwarf
with a very low mass ($M=0.17\ M_\odot$) and a temperature of $11\,080$ K.
Table~\ref{tbl_par} summarizes the properties of LP~400-22. 
Since white dwarfs with masses below $0.4\ M_\odot$ 
must have been formed in close binary systems, radial velocity
measurements and infrared photometry are required to determine whether
LP~400-22 has a close companion. On the other hand, a lack of
radial velocity variations would indicate that LP~400-22 lost
its close massive companion following a type Ia supernova event.

\acknowledgments

This research is supported in part by a NASA/GALEX grant (NNG05GE33G).
A.K. is supported by GA \v{C}R 205/05/P186.
T.D.O. acknowledges support from the NSF (AST 0206115).
J.A.S. was supported on a NASA GSRP Training Grant, NGT-51086.
N.M.S. acknowledges support from a NASA GSRP grant (NST 200415) and
NSF AST 02-05875.
We thank Paul Hintzen for acquiring the original KPNO spectrum of LP~400-22.
This research has made use of the VizieR catalogue access tool, CDS, 
Strasbourg, France.
Based on observations obtained with the APO
3.5 m telescope, which is owned and operated by the
Astrophysical Research Consortium.

\end{document}